\documentstyle[sprocl]{article}

\input epsf

\begin{document}
\title{ACCURATELY DETERMINING INFLATIONARY PERTURBATIONS}
\author{ ANDREW R. LIDDLE and IAN J. GRIVELL }
\address{Astronomy Centre, University of Sussex,\\ Brighton BN1 9QH, Great 
Britain}
\maketitle
\abstracts{Cosmic microwave anisotropy satellites promise extremely accurate 
measures of the amplitude of perturbations in the universe. We use a 
numerical code to test the accuracy of existing approximate expressions for 
the amplitude of perturbations produced by single-field inflation models. We 
find that the second-order Stewart--Lyth calculation gives extremely 
accurate results, typically better than one percent. We use our code to 
carry out an expansion about the general power-law inflation solution, 
providing a fitting function giving results of even higher accuracy.}

\section{Motivation}

Two newly approved satellites which will observe microwave background 
aniso\-tropies, MAP and COBRAS/SAMBA, promise to measure the amplitude of 
these aniso\-tropies at the percent level. Recent theoretical work has 
demonstrated that these anisotropies can be predicted from a given spectrum 
of adiabatic perturbations, to an accuracy of one percent or 
better.\cite{CMB} The most promising theory for the generation of adiabatic 
perturbations is cosmological inflation, and here we address the question of 
whether or not it is possible to predict the spectrum from an inflationary 
model at a similar level of accuracy, in order to allow one to take full 
advantage of forthcoming observations. Full details can be found in Ref.~2, 
which also discusses gravitational waves.

We shall only consider models with a single scalar field $\phi$, moving in 
an arbitrary potential $V(\phi)$. This is not to say that accurate 
calculations cannot be done in more general models, but there typically they 
have to be done on a case-by-case basis, whereas the single-field models can 
be analyzed simultaneously. Our aim is to compute the amplitude ${\cal 
P}_{{\cal R}}(k)$ of the curvature perturbation ${\cal R}$ on a given fixed 
comoving scale $k$, where the precise terminology is defined in Ref.~2. From 
this one could also compute the spectral index, its rate of change, etc. 
Typically though the corrections we discuss to the standard results are 
observationally negligible except for the amplitude itself.

\section{Framework}

The best calculational framework is that introduced by Mukhanov~\cite{M} and 
exploited by Lyth and Stewart.\cite{LS,SL} The only knowledge we require of 
the background (homogeneous) evolution is of the combination $z = a 
\dot{\phi}/H$, where $a$ is the scale factor and $H$ the Hubble parameter. 
The perturbation can be expressed by a gauge-invariant potential $u = -z 
{\cal R}$. Finally, we use conformal time $\tau$. Then a fourier mode of the 
perturbation $u$ obeys the equation 
\begin{equation}
\frac{d^2 u_k}{d\tau^2} + \left( k^2 - \frac{1}{z} \frac{d^2 z}{d\tau^2}
	\right) u_k = 0
\end{equation}

Typically we don't know the background evolution $z(\tau)$, so we don't even 
know the equation we are trying to solve. Fortunately though it possesses 
two asymptotic regimes where the solution can be found:
\begin{eqnarray}
& u_k = \frac{1}{\sqrt{2k}} e^{-ik\tau} \quad \quad & k \gg aH \\
& u_k \propto z & k \ll aH
\end{eqnarray}
The former is the flat-space limit and includes the appropriate quantum 
normalization. The latter is the growing mode, and corresponds to ${\cal 
R}_k$ constant. Both regimes are independent of $z(\tau)$, so the amplitude 
of perturbations only depends on the transition regime. We can therefore 
expand $z(\tau)$ about the time $\tau$ where $k = aH$.

This expansion of the background evolution can be carried out using the 
slow-roll expansion.\cite{LPB} Inflation is described using as a fundamental 
quantity the Hubble parameter as a function of the scalar field value, 
$H(\phi)$. From its derivatives one defines a series of slow-roll 
parameters, the first two being 
\begin{equation}
\epsilon = \frac{m_{{\rm Pl}}^2}{4\pi} \left( \frac{dH/d\phi}{H} \right)^2
	\quad ; \quad \eta = \frac{m_{{\rm Pl}}^2}{4\pi} 
\frac{d^2
	H/d\phi^2}{H}
\end{equation}
The slow-roll approximation demands that all these are small. They can 
readily be computed for a given $V(\phi)$.
 
\section{Results}

We solve the mode equation numerically, in order to compare with existing 
calculations. Figure 1 shows four panels, testing the standard slow-roll 
approximation result ${\cal P}_{\cal R}^{1/2} = H^2/2\pi|\dot{\phi}|$, the 
improved second-order result of Stewart and Lyth,\cite{SL} an approximation 
based on using the exact power-law inflation result,\cite{LS,GL} and finally 
a new expression we derived~\cite{GL} by expanding about a general power-law 
solution.

\begin{figure}[t]
\centering 
\leavevmode\epsfysize=8.3cm \epsfbox{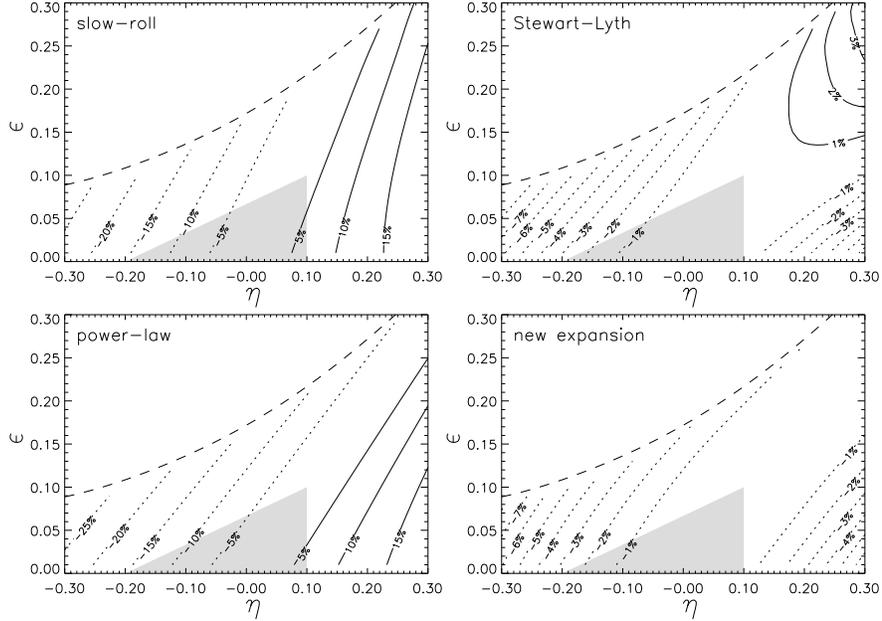}\\ 
\caption[contours]{A comparison of four analytic expressions with the exact 
numerical result, with the contours showing the relative error. Each point 
in the $\epsilon$--$\eta$ plane corresponds to a different inflation model. 
Above the dashed line is a forbidden region (inflation ends too quickly), 
while the shaded area indicates those models currently preferred by 
observational data constraining the slope of the spectrum and amount of 
gravitational waves.} 
\end{figure} 

We conclude that it is possible to obtain very accurate results numerically, 
and also that anyway the Stewart--Lyth calculation~\cite{SL} is already 
extremely accurate, being better than two percent accurate in all the 
parameter region currently favoured by observations.

\section*{Acknowledgments} 

The authors were supported by the Royal Society.


\section*{References}
\begin{thebibliography}{99} 
\bibitem{CMB} W Hu, D Scott, N Sugiyama and M White, {\it Phys. Rev.} D
	{\bf 52}, 5498 (1995); U Seljak and M Zaldarriaga, {\it Astrophys.
	J.} {\bf 469}, 437 (1996).
\bibitem{GL} I J Grivell and A R Liddle, {\it Phys. Rev.} D {\bf 54}, 7191 
	(1996).
\bibitem{M} V F Mukhanov, {\it Pis'ma Zh. Eksp. Teor. Fiz.} {\bf 41}, 402
	(1985), {\it Zh. Eksp. Teor. Fiz.} {\bf 94}, 1 (1988).
\bibitem{LS} D H Lyth and E D Stewart, {\it Phys. Lett.} B{\bf 274}, 168
	(1992).
\bibitem{SL} E D Stewart and D H Lyth, {\it Phys. Lett.} B{\bf 302}, 171
	(1993).
\bibitem{LPB} A R Liddle, P Parsons and J D Barrow, {\it Phys. Rev.} D 
	{\bf 50}, 7222 (1994).
\end{thebibliography}
\end{document}